\documentclass{article}

\usepackage{amsmath}
\usepackage{amssymb}
\newtheorem{definition}{Definition}

\newcommand{\IO}{\ensuremath{\diamond}}
\newcommand{\IN}{\ensuremath{\triangleleft}}
\newcommand{\OUT}{\ensuremath{\triangleright}}
\newcommand{\HB}{\hfill{$\square$}}
\newcommand{\U}{\ensuremath{\:\cup\:}}
\newcommand{\INT}{\ensuremath{\varepsilon}}

\begin{document}
\title{A Trace Logic for Local Security Properties\thanks{We would like to thank Cabernet and the EYES Project (IST- 2001-34734) for their support of this work.}}  
\author{Ricardo Corin, Sandro Etalle, Pieter Hartel\\
Faculty of Computer Science,\\
University of Twente,  The Netherlands\\
Email: \texttt{\{corin,etalle,pieter\}@cs.utwente.nl}\\
\ \\
Antonio Durante\\
Universit\`a di Roma ``La Sapienza'', 
Italy\\
Email: \texttt{durante@dsi.uniroma1.it}
}

\maketitle


\begin{abstract} We propose a new simple \emph{trace} logic
  that can be used to specify \emph{local security properties}, i.e.
  security properties that refer to a single participant of the
  protocol specification.  Our technique allows a protocol designer to
  provide a formal specification of the desired security properties, and
  integrate it naturally into the design process of cryptographic
  protocols.  Furthermore, the logic can be used for formal
  verification. We illustrate the utility of our technique by exposing
  new attacks on the well studied protocol TMN.\\
{\bf Revision history: Nov 5 2004. Comments: fixed typos.}\\
{\bf Revision history: Nov 30 2004. Comments: added (variable) events Definition~\ref{sec:trace-logic} , update bindings in execution, Definition \ref{procedure}, removed Implementation section.}
\end{abstract}
\section{Introduction}

Cryptographic protocols are typically designed to meet security goals
such as authentication and confidential key exchange. These goals,
usually called \emph{security properties}, can be correctly
accomplished if some of the values exchanged during the protocol run
satisfy, for instance, classical properties like authenticity,
confidentiality, or freshness.

Often, the specification of security properties is carried out by
writing of ``global'' security properties. These security
specifications do not depend from any principal's point of view. Thus,
to refer to a specific principal, global security properties are
usually defined using extra protocol events \cite{Lowe97,Mea96}.

In this paper, on the other hand, we propose a logic that can be used
to express \emph{local security properties}, i.e. properties that
refer to the specification of \emph{one} agent, namely the agent which
they belong to. As we show in the following sections, local security
properties are expressive enough to assert the properties that are
commonly desired for cryptographic protocols (e.g., freshness of a
nonce.)

The advantage of local properties is that they allow a designer to
 specify the security properties that should hold,
according to each participant, at each protocol execution point.  For
instance, a property like freshness of a nonce can be specified as a
formula that is connected directly to the corresponding participant who
receives that nonce. Furthermore, since these formulae correspond to
each principal, they depend \emph{only} on information of that
principal, as opposed to a global formula that can depend on the whole
network state. Thus, a local formula can be bound to each principal
and then be ``plugged in'' into any other network specification. This
enables potential composability of the specifications.

Consequently, using local properties, it is possible to
integrate the specification of the (logical) security properties that
a protocol has to meet \emph{within} the (algorithmic) specification of
the protocol itself. This yields an integrated technique for protocol
engineering that combines tightly the design and the analysis phase,
resulting in a shorter design-verification feedback loop.

We illustrate our approach by studying the TMN protocol \cite{tmn90} for
which we have found two new attacks.

\emph{Plan of the paper.} In Section \ref{sec:protmodel}, we describe
our security protocol model. Then, in Section \ref{sec:tracemodel} we
introduce our trace logic language.  In Section \ref{sec:tmn}, the
TMN protocol is studied and some novel attacks upon it are
presented. In Section \ref{sec:relatedwork} elaborates the related
work and finally conclusions and future work are discussed in Section
\ref{sec:conclusions}.

\section{Protocol Model}
\label{sec:protmodel}

A protocol step is usually specified using the standard notation $A
\rightarrow B: M$. Here, $M$ is a message built from:
\begin{itemize}
\item atomic terms, that is constants (written in lowercase) and
  variables (which are capitalized). Constants may be nonces (e.g.
  \emph{na}) or agent identities (e.g. $a$). A special constant \INT \ 
  denotes the intruder.
\item constructed terms, that is a finite application of operators
\emph{encryption} $M_K$, \emph{pairing} $M_1,M_2$, \emph{hashing}
$h(M)$ and finally \emph{public key} $pk(M)$ over atomic terms.
\end{itemize}
However, the $A \rightarrow B: M$ notation is unsuitable for formal
verification.  In fact, in a protocol step, two different events take
place: $A$ sends message $M$, and $B$ receives message $M'$.  In
presence of an intruder, $M$ might not be equal to $M'$.  Moreover,
not even the identities of the correspondent communication parties may
be the same (i.e., $A$ sends to $B'$ and $B$ receives from $A'$.)  It
is therefore convenient to take an approach that considers separately
each agent's point of view; this is the idea of \emph{protocol roles}.

\begin{definition}
  A \emph{protocol role} is a pair $\langle A , [M_1 \IO B_1, ..., M_n
  \IO B_n] \rangle$, where $A, B_1,...,B_n$ are variables, $\IO \in
  \{ \IN, \OUT \}$ and $M_1,...,M_n$ are messages. \HB
\end{definition}

Given a protocol role $\langle A , [M_1 \IO B_1, ..., M_n \IO B_n]
\rangle$, $A$ is called the \emph{identity} of the role, while elements 
$M_i \IO B_i, i=1..n$ 
are the \emph{actions} of the role: $M \triangleright B$ is a \emph{send} 
action, while $M \triangleleft B$ is a \emph{receive} action.

Protocol roles in a security protocol often receive (self explanatory)
names such as \emph{initiator, responder} and \emph{server}.  For
example,
\begin{equation}
\label{eq}
 \emph{responder(A,B,Na)}=\langle B, [pk(\emph{Na}) \triangleleft A]\rangle
\end{equation}
defines a responder role in which there is only one action, the receipt of
\emph{Na} from $A$.

Notice that in (\ref{eq}), the variables $A,B,Na$ are still
uninstantiated (we borrow this concept from logic programming: as long
as no value is assigned to a variable, we call it
\emph{uninstantiated}, and \emph{instantiated} otherwise.) In fact, a
protocol role is \emph{parametric}, thus representing a template.  By
appropriately (partially) instantiating a finite number of protocol
roles, a \emph{system scenario} can be obtained:
\begin{definition}
A system scenario is a multiset of (partially) intantiated protocol roles.
\end{definition}

Typically, a system scenario determines how many sessions are present
and which agents play which roles. For instance, the system scenario
$$\{ \emph{responder(A,b,Na), responder(C,d,Nc)} \} $$ (where
\emph{responder} is the role defined above) defines a system scenario
with two responders (notice that there are no corresponding
\emph{initiators}), one played by $b$ and the other by
$d$. Uninstantiated variables represent unknown values: for example,
variable $A$ in the first responder role represents the (unknown)
communicating party of $b$.

\subsection{Trace Semantics}

Executions of system scenarios are described using \emph{traces}, which 
are in turn composed of \emph{events}, i.e. single actions performed 
by an agent.

\begin{definition} An \emph{event} is a pair $\langle A : M \IO  B \rangle$
  where $A,B$ are agent's names, $\IO \in \{ \IN, \OUT \}$ and $M$ is
  a message.\HB
\end{definition}
The event $\langle A : M \triangleright B \rangle$ should be read as
``agent $A$ sends message $M$ with \emph{intended destination} $B$''.
On the other hand, $\langle B : M \triangleleft A \rangle$ stands for
``agent $B$ receives message $M$ \emph{apparently from} $A$''.

To analyze the protocol, we combine the system scenario with the usual
Dolev-Yao intruder \cite{DY83}, who can perform the usual actions: 
intercept and learn any sent message, store the information
contained in intercepted messages for later use, and introduce into
the system new messages forged using information the intruder knows. The
information obtained by the intruder is stored in a set of terms $K$
called the \emph{intruder's knowledge}\footnote{Because of the symbolic nature of the analyzer, in practice an event can contain variables, which stand for something the intruder can generate (see \cite{CE02} for details.)
}

Now we are ready to describe the execution of a system scenario, represented
by the notion of a \emph{run}.  

\begin{definition}
\label{procedure}
Let $S$ be a system scenario, and $K$ be the intruder's initial
knowledge, consisting of constants representing agents identities and
their public keys.  Let $tr$ be an initial empty trace.  A \emph{run} of $S$
is a trace obtained by a reiterated sequence of the following steps:
\begin{enumerate}
\item a non-empty role in $S$ is chosen nondeterministically, and   
  its first action $p$ is removed from it. Let $a$ be the identity of 
the chosen role.
\item if $p = t \triangleright y$, then:
 \begin{enumerate}
  \item $t$ is added to the knowledge of the intruder,
  $K := K \U \{t\}$
  \item event $e = \langle a : p \rangle$ is added to $tr$, $tr := tr 
\cdot e$
\end{enumerate}

\item if $p =  t \triangleleft y $, then:
\begin{enumerate}
\item it is checked if the intruder \INT\ can generate $t$ using the
knowledge $K$\footnote{We adopt Millen and Shmatikov's constraint solving procedure
\cite{MS01} for checking if the intruder can generate a term $t$ using
knowledge $K$. This procedure may involve instantiation of variables
in $t$ or $K$; for example, $t$ may unify with a term in $K$,
representing that $t$ is already in $K$, i.e., is already known by the
intruder (see \cite{MS01} for details.)}, if so, then event $e = \langle a : p
\rangle$ is added to the trace: $tr := tr \cdot e $.

\item If \INT\ cannot generate such a message, then the run stops.
\end{enumerate}\HB
\end{enumerate}
\end{definition}

\section{A Trace Logic}
\label{sec:tracemodel}

In this section we introduce a trace logic language for defining  
local security properties.

\begin{definition} 
\label{sec:trace-logic}
A trace logic formula is generated according to the following grammar:\\

\begin{tabular}{c l}

$F ::=$ & $\begin{tt}{true}\end{tt} \ | \ \begin{tt}{false}\end{tt} \ | \ F_1 \wedge F_2 \ | \ F_1 \vee F_2 \ | \ F_1 \rightarrow F_2 \ | \ \forall e \in tr \ : F $\\ 
&$\ | \ \exists e \in tr\ : F \ | \exists t : F \ | \neg F \ | e_1 = e_2 \ | \ e_1 \preceq e_2 $\\
\end{tabular}\\

where $e$, $e_1$ and $e_2$ are (variable) events. \HB
\end{definition} 

The conjunction of two formulae has the usual significance: $F_1
\wedge F_2$ is \emph{true} if both $F_1$ and $F_2$ are \emph{true};
the disjunction operator $\vee$ and implication $\rightarrow$ are
analogous.  On the other hand, the meaning of constructors $\forall e
\in tr\ : F$ and $\exists e \in tr\ : F$ is non-standard.  Since a
trace formula is going to be evaluated on a certain input trace,
constructors $\forall$ and $\exists$ allow us to reason about the
events in the input trace: $\forall e \in tr\ : F$ asserts that every
event $e$ in the input trace satisfies formula $F$, while $\exists e
\in tr\ : F$ express that some event in the input trace satisfies
formula $F$.  Notice that $tr$ is not a variable, it is just part of
the operators name to emphasize that $e$ ranges over the system trace.
Even though this gives a ``temporal'' flavor to our logic, we
anticipate that these constructors only operate on \emph{past} events,
recorded in the input trace (see later). Formula $\neg F$ has the
usual meaning of negation. Differently from the above operators,
$\exists t: F$ quantifies $t$ over all messages and agents space.
Finally, predicates $e_1 = e_2$ and $e_1 \preceq e_2$ allow us to
compare events: the former asserts equality, and the latter
\emph{subterm} inclusion.

While the choice of these constructors may seem rather \emph{ad hoc}
for our purposes, we believe this logic can in fact be quite
expressive, and allow us to assert a fairly large set of interesting
security properties, as will be shown later.

Next, we define the precise meaning of a
trace logic formula.

\begin{definition}
Let $\cal{F}$ be the set of well-formed trace logic formulae, and
\emph{TR} be the set of traces, then the semantic
function $[\![ \cdot ]\!]\cdot :\begin{cal}{F}\end{cal} \times
\emph{TR} \rightarrow \{true, false\}$ is defined as follows:\\[0.5mm]
\begin{small}
\begin{tabular}{l l}
$[\![ \begin{tt}{true}\end{tt} ]\!]\ tr $&  = $ true $\\
$[\![ \begin{tt}{false}\end{tt} ]\!]\ tr $& $ = false $\\
$[\![  F_1 \wedge F_2 ]\!]\ tr $& $ = true \emph{ iff } [\![ F_1 ]\!]\ tr = [\![ F_2 ]\!] \ tr =true $\\
$[\![  F_1 \vee F_2 ]\!]\ tr $& $ = true \emph{ iff } [\![ F_1 ]\!]\ tr = true \emph{ or } [\![ F_2 ]\!] \ tr =true $\\

$[\![  F_1 \rightarrow F_2 ]\!] \ tr $&$ = true \emph{ iff } [\![ F_1 ]\!] \ tr
\emph{ implies } [\![ F_2 ]\!] \ tr$\\
$[\![  \forall e \in tr\ : F  ]\!] \ tr $& $ = $true \emph{ iff,  for each event} $x$ of $tr$, $[\![ F[^x / e] ]\!] \ tr =true $ \\

$[\![  \exists e \in tr\ : F  ]\!] \ tr $& $ = $true \emph{ iff,  for some event} $x$ of $tr$, $[\![ F[^x / e] ]\!]\ tr =true $\\

$[\![  \exists t\ : F  ]\!] \ tr $& $ = $true \emph{ iff,  for some message or agent} $x$, $[\![ F[^x / t] ]\!]\ tr =true $\\

$[\![  \neg F ]\!]\ tr $& $ = true \emph{ iff } [\![ F ]\!]\ tr=false $\\

$[\![  e_1 = e_2  ]\!] \ tr $& $ = true \emph{ iff event $e_1$ is equal to 
event $e_2$}$ \\ 
$[\![  t_1 \preceq t_2  ]\!] \ tr $& $ = true $\emph{ iff, if }$t_1$ \emph{is a subterm of} $t_2$ 
\end{tabular}
\end{small}
\end{definition}\HB

Here, $F[x / y]$ is the result of substituting each occurrence of $y$ with $x$ in $F$.

For the sake of notation's simplicity, we assume that all variables
that are not explicitly quantified are \emph{existentially} quantified
(over the set of messages and agents).  This simplifies the notation
considerably.

In the future, we plan to endow our logic with a proof system that
allow us to relate proofs of formulae with the intended meaning given
by $[\![ \cdot ]\!]\cdot$. In the present work, we are more interested
in exploring the expressive power of security specifications; We plan
to continue this work by addressing the issue of using our logic for
automatic formal verification.

\subsection{Appending local security properties to protocol roles}
Now, we are ready to combine the definition of protocol roles and local security properties
 to obtain \emph{extended protocol roles} and \emph{extended
  system scenarios}. Intuitively, the idea is to embed the logical
security properties within the protocol specification.

\begin{definition}
  An \emph{extended protocol role} is a triple $\langle A , [M_1 \IO
  B_1 : F_1, ..., M_n \IO B_n : F_n ] \rangle$, where $\{A,
  B_1, ..., B_n\} \subset Var$, $M_1,...,M_n$ are messages, $\IO \in
  \{ \IN, \OUT \}$ and $F_1,...,F_n$ are trace logic
  formulae.\HB
\end{definition}

Intuitively, adding a formula $F_i$ after a protocol role action means
that $F_i$ must hold after the execution of the action.  Notice that
instantiation of an extended protocol role also affects the variables
of an attached local security property. This formalizes the notion of
a security property being `local', that is a security specification
that takes into account the principal's point of view.  Also, $F_i$ is
going to be evaluated w.r.t. the system trace, which contains the
events \emph{up to} at that precise execution time. This, as we already
mentioned, illustrates the ``past flavour'' nature of our formulae.

Similarly, we can define an extended system scenario as a multiset of
(partially instantiated) extended protocol roles. 

\subsection{Verifying the local security properties}
\label{verification}

To evaluate the local security properties, we extend the Definition
\ref{procedure} to the extended system scenarios introduced in last
section:

\begin{definition}
\label{procedure}
Let $S$ be an \emph{extended} system scenario, and $K$ be the intruder's initial
knowledge, consisting of constants representing agents identities and
their public keys.  Let $tr$ be an initial empty trace.  A \emph{run} of $S$
is a trace obtained by a reiterated sequence of the following steps:
\begin{enumerate}
\item a non-empty role in $S$ is chosen nondeterministically, and   
  its first action $p$ is removed from it. Let $a$ be the identity of 
the chosen role.

\item if $p = t \triangleright y : F$, then:
 \begin{enumerate}
  \item if $[\![F]\!]tr$ holds, then update the resulting bindings (which appear from the existencially quantified variables) and continue. Otherwise, the run stops.
  \item $t$ is added to the knowledge of the intruder $\varepsilon$,
  $K := K \U \{t\}$
  \item event $e = \langle a : p \rangle$ is added to $tr$, $tr := tr 
\cdot e$
\end{enumerate}

\item if $p =  t \triangleleft y :F $, then:
\begin{enumerate}
\item it is checked if the intruder \INT\ can generate $t$ using the
knowledge $K$ (see below), if so, then event $e = \langle a : p
\rangle$ is added to the trace: $tr := tr \cdot e $.
\item if $[\![F]\!]tr$ holds, then continue. Otherwise, the run stops.

\item If \INT\ cannot generate such a message, or \INT\ simply decides to finish the execution, then the run stops.
\end{enumerate}\HB
\end{enumerate}
\end{definition}

For example, consider the role:
$$\emph{responder(B, A, Na)} = \langle B, [ \emph{Na} \triangleleft A:
F]\rangle$$ where $F = \exists
e: e=\langle A: \emph{Na} \triangleright B\rangle$.

After the responder $B$ receives the nonce \emph{Na}, $F$ checks that 
$A$ had sent \emph{Na} to $B$ before. Now, consider the
singleton scenario $\{ \emph{responder(b,A,Na)} \}$. In this scenario,
there is only one honest responder role, played by $b$. Now, suppose
this responder role receives, from the intruder $\varepsilon$, a nonce
$ni$ as \emph{Na}. Therefore, according to Definition \ref{procedure},
we have trace $tr= \langle \INT: ni \triangleright b \rangle$. The
next step involves evaluation of $[\![F]\!] tr$ to see if the local
security property $F$ holds: clearly, we can see that $[\![ \exists e:
e=\langle A: \emph{Na} \triangleright b\rangle ]\!] \langle \INT: ni
\triangleright b \rangle$ evaluates to \emph{true}, assigning \INT\ to $A$ and $ni$ to \emph{Na}. 


\section{A Case Study: the TMN protocol}
\label{sec:tmn}
We apply our technique to a well known case study, the TMN protocol
\cite{tmn90}. This protocol has been thoroughly studied, see for
example \cite{Roscoe96,MMS97,LR97}. However, in this section we
present some vulnerabilities that we believe no one
has noticed before.

\subsection{Original Version}
The original version of TMN was proposed
for achieving key distribution between two users:\\

\begin{tabular}{c l}
Message $1$.&$A \rightarrow S \ : \ A,S,B, \{\emph{R}_1\}_{pk(S)} $ \\
Message $2$.&$S \rightarrow B \ : \ S,B,A $ \\
Message $3$.&$B \rightarrow S \ : \ B,S,A, \{\emph{R}_2\}_{pk(S)} $ \\
Message $4$.&$S \rightarrow A \ : \ S,A,B, v(\emph{R}_1,\emph{R}_2) $ \\
\end{tabular}\\

We denote Vernam
encryption by $v(t_1,t_2)$\footnote{We currently model Vernam
encryption as normal symmetric encryption, and not as full exclusive
$\mathtt{xor}$.}.
Here, keys $R_1$ and $R_2$ are sent from $A$ and $B$ to $S$,
respectively. After Message 4 is received, $A$ can obtain $R_2$, thus
making $R_2$ the shared key between $A$ and $B$.

\subsubsection{TMN protocol roles.}
The first step in our design and verification technique is to obtain
the protocol roles from the standard notation:
\begin{itemize}
\item Initiator: 
$\langle A, [ A,S,B, \{\emph{R}_1\}_{pk(S)} \triangleright S : F_1, \ \ 
S,A,B,v(\emph{R}_1,\emph{R}_2) \triangleleft S:F_2 ] \rangle$

\item Responder: 
$\langle B, [ S,B,A \triangleleft \ S:F_3, \ \ 
B,S,A, \{\emph{R}_2\}_{pk(S)} \triangleright S: F_4 ] \rangle$

\item Server: 
$\langle S, [ A,S,B, \{\emph{R}_1\}_{pk(S)} \triangleleft A : F_5, \ \ 
S,B,A \triangleright B : F_6, \ \ B,S,A, \{\emph{R}_2\}_{pk(S)} 
\triangleleft B: F_7, \ \ 
S,A,B, v(\emph{R}_1,\emph{R}_2) \triangleright A: F_8 ] 
\rangle$
\end{itemize}

This translation can be tedious and error-prone when protocols get
large; however, we believe this step can be mostly automated (eg. by a
tool assisting the user.)

The original version of TMN suffers from several secrecy attacks over
$R_2$ above, as exposed for instance in \cite{LR97}. Thus, we will concentrate on two modified versions of the protocol.

\subsection{First modification} A
replay attack against TMN was exposed by Simmons \cite{SIM94}. The
attack exploits the fact that the messages to the server from $A$ and
$B$ (Message 1 and Message 3) can be replayed.  To solve this
defficiency, Tatebayashi and Matsuzaki introduce timestamps
in messages 1 and 3 \cite{tmn90}:\\

\begin{tabular}{c l}
Message $1$.&$A \rightarrow S \ : \ A,S,B, \{\emph{T}_A, \emph{R}_1\}_{pk(S)} $ \\
Message $2$.&$S \rightarrow B \ : \ S,B,A $ \\
Message $3$.&$B \rightarrow S \ : \ B,S,A, \{\emph{T}_B, \emph{R}_2\}_{pk(S)} $ \\
Message $4$.&$S \rightarrow A \ : \ S,A,B, v(\emph{R}_1,\emph{R}_2) $ \\
\end{tabular}\\

In this new protocol, after receiving $T_A$ and $T_B$, the server can
check for the timeliness of these timestamps.  According to
Tatebayashi and Matsuzaki, this new protocol version guarantess the
freshness of $R_1$ and $R_2$. To check if this is true, we can specify
the freshness requirements of $R_1$ and $R_2$ as a local security
properties of server $S$:
$$Fresh_{R_i} = \forall e \in tr: last\_event(e) \ \vee \ 
\neg(\emph{R}_i \preceq msg(e) ) \text{ (for }i=1,2)$$
Where primitive $msg(\cdot)$ projects the message of an event, defined
as $msg(\langle x: m \IO y\rangle)=m$ and predicate $last\_event(e)$
is a primitive that is true iff $e$ is the last event of trace $tr$.
The definition of this primitive is straightforward: $[\![
last\_event(e) ]\!] \ tr = true$ iff $tr=tr'\cdot e$.
$Fresh_{R_1}$ and $Fresh_{R_2}$ are expressing that 
 $\emph{R}_1$ and $\emph{R}_2$, respectively, are fresh.

The last step involves deciding where to put $Fresh_{R_1}$ and
$Fresh_{R_2}$ in the server role. This is easy: we make the decision that the formulae for checking the freshness of the received values should be placed \emph{as soon as the values are received}. Thus, $Fresh_{R_1}$ can be
put as $F_5$, that is, after $R_1$ is received. Similarly, we set $Fresh_{R_2}$ as $F_7$.

\subsection{First novel attack}
After verification, we found a violation of formula $F_5$ 
(that is, freshness of $R_1$). The attack is reported in Table \ref{es_2}.
\begin{table} [h]
\begin{center}
\caption{\label{es_2}  $R_1$ freshness attack. $\varepsilon(s)$ is $\varepsilon$ masquerading as $s$.  $\alpha$ and $\beta$ denote two different runs.}
\begin{tabular}{l l} \hline 
Message $\alpha.1 $.&$ a \rightarrow \INT(s) \ : \ a,s,b, \{\emph{t}_a, \emph{r}_1\}_{pk(s)} $ \\
Message $\alpha.1'$.&$ \INT(a) \rightarrow s \ : \ a,s,b, \{\emph{t}_{e1}, \emph{r}_e\}_{pk(s)} $ \\
Message $\alpha.2 $.&$ s \rightarrow b \ : \ s,b,a $ \\
Message $\alpha.3 $.&$ b \rightarrow \INT(s) \ : \ b,s,a, \{\emph{t}_b, \emph{r}_2\}_{pk(s)} $ \\
Message $\alpha.3'$.&$ \INT(b) \rightarrow s \ : \ b,s,a, \{\emph{t}_a, \emph{r}_1\}_{pk(s)} $ \\
Message $\alpha.4 $.&$ s \rightarrow \INT(a) \ : \ s,a,b, v(\emph{r}_e,\emph{r}_1) $ \\
Message $\beta.1$.&$ \INT(a) \rightarrow s \ : \ a,s,b, \{\emph{t}_{e2}, \emph{r}_1\}_{pk(s)} $ \\
\hline
\end{tabular}\\
\end{center}
\end{table}

In this attack, the intruder starts replacing messages $\alpha.1$ with
$\alpha.1'$ and $\alpha.3$ with $\alpha.3'$, and finally obtains $r_1$
from message $\alpha.4$. But, when it wants to use it in a new run
$\beta$, even if the intruder uses a new (not expired) timestamp $t_{e2}$, the
formula $F_5$ does not hold since $r_1$ is not fresh (note that $s$ is
the \emph{same} server, involved in both runs $\alpha$ and $\beta$).
It is important to notice why this attack represents a
vulnerability of the protocol. According to Tatebayashi and Matsuzaki,
the server has to check for the validity of the timestamps in order
to guarantee the freshness of $R_1$ and $R_2$; as we can see in this
attack, this is not sufficient.  To the best of our knowledge, this
vulnerability was never exposed before.\\

\subsection{Second modification} A modification to assure
authentication of the initiator and responder to the server consists in
using $S_A$ and $S_B$, shared secrets between $S$ and $A$ and $B$
respectively, in the following manner:\\

\begin{tabular}{c l}
Message $1$.&$A \rightarrow S \ : \ A,S,B, \{\emph{T}_A, \emph{S}_A, \emph{R}_1\}_{pk(S)} $ \\
Message $2$.&$S \rightarrow B \ : \ S,B,A $ \\
Message $3$.&$B \rightarrow S \ : \ B,S,A, \{\emph{T}_B, \emph{S}_B, \emph{R}_2\}_{pk(S)} $ \\
Message $4$.&$S \rightarrow A \ : \ S,A,B, v(\emph{R}_1,\emph{R}_2) $ \\
\end{tabular}\\

After receiving messages 1 and 3, the server can authenticate $A$ and
$B$, respectively, since (by assumption) secrets $S_A$ and $S_B$ are
shared \emph{only} between the server and the respective agents.  To
check if the protocol accomplishes the authentication goal of $A$ and
$B$ to $S$, we translate this in a formula that states that if $S$
received a message $M$ apparently from $A$ (resp. $B$), then it was
\emph{really} sent by $A$ ($B$).  
%
Server $S$ authenticates $A$ after receiving the first message, 
so at that point we set our formula: 
$F_5 =  \exists e: e= \langle A: \ A,S,B, \{\emph{T}_A, \emph{S}_A, \emph{R}_1\}_{pk(S)} \triangleright S  \rangle$. Similarly, $S$ authenticates $B$ after the third message: $F_7 =  \exists e: e= \langle B: \ B,S,A, \{\emph{T}_B, \emph{S}_B, \emph{R}_2\}_{pk(S)} \triangleright S  
\rangle $.

We performed verification with some test scenarios and did not find any trace that violates the above 
security requirements. Thus, we can regard the protocol to be secure for the 
system scenarios we tested; of course, bigger scenarios can be tested to 
increase confidence about the protocol security.

\subsection{Mutual authentication} Even though Tatebayashi and Matsuzaki do not state the
mutual authentication of $A$ and $B$, it is interesting to consider
this case (Lowe and Roscoe \cite{LR97} also discuss this.) We can
translate this requirement by redefining two formulae, namely $F_3$
and $F_2$. We define $F_3$ to express the local security property of A
to B and $F_2$ expressing the authentication of B to A:

\begin{itemize}
 \item $M$ authenticity of $A$ to $B$: $ F_3 = \exists e: e=\langle A:
 A,S,B,\{T_A,S_A,R_1\}_{pk(S)} \triangleright S  \rangle$;
 \item  $M$ authenticity of $B$ to $A$: $ F_2 = \exists e: e= \langle B:  B,S,A,\{T_B,
 S_B,R_2\}_{pk(S)} \triangleright S  \rangle $.
 \end{itemize}

Proceeding with verification, we found traces that violate $F_2$ and
$F_3$.  The attack trace for $F_3$ is straightforward, consisting in
only one message, sent from $\INT(s)$ to $b$: $s,b,a$. But this is
sufficient to violate formula $F_3$, since when $b$ receives
$s,b,a$ she wants to check if $a$ sent $
a,s,b,\{t_a,s_a,r_1\}_{pk(s)}$, which she did not (this attack is similar
to attack 7.1 in \cite{LR97}.)

\subsection{Novel authentication attacks}
In Table \ref{es_3} we report two attacks that violate $F_2$.

\begin{table} [h]
\begin{center}
\caption{\label{es_3}  B to A authentication attacks}
\begin{tabular}{|c l|l |} \hline
$\alpha.1$.&$ a \rightarrow \INT(s) \ : \ a,s,b, \{\emph{t}_a, \emph{s}_a,
\emph{r}_1\}_{pk(s)} $
& $\alpha.1$.$ a \rightarrow \INT(s) \ : \ a,s,b, \{\emph{t}_a,
\emph{s}_a,\emph{r}_1\}_{pk(s)} $ \\

$\alpha.2$.&$ \INT(s) \rightarrow b \ : \ s,b,\INT $
&$\beta.1$.$ \INT \rightarrow s \ : \ \INT,s,a, \{\emph{t}_e,
\emph{s}_e,
\emph{re}_1\}_{pk(s)}  $ \\

$\alpha.3$.&$ b \rightarrow \INT(s) \ : \ b,s,\INT, \{\emph{t}_b,
\emph{s}_b, \emph{r}_2\}_{pk(s)} $
& $\beta.2$.$ s \rightarrow \INT(a) \ : \ s,a,\INT $ \\

$\beta.1$.& $ \INT(a) \rightarrow s \ : \ a,s,b, \{\emph{t}_a, \emph{s}_a,
\emph{r}_1\}_{pk(s)} $
&$\beta.3$.$ \INT(a) \rightarrow s \ : \ a,s,\INT, \{\emph{t}_a,
\emph{s}_a,
\emph{r}_1\}_{pk(s)}$ \\

$\beta.2$.&$ s \rightarrow \INT(b) \ : \ s,b,a $
& $\beta.4$.$ s \rightarrow \INT \ : \ s,\INT,a,
v(\emph{re}_1,\emph{r}_1) $ \\

$\beta.3$.&$ \INT(b) \rightarrow s \ : \ b,s,a, \{\emph{t}_b, \emph{s}_b,
\emph{r}_2\}_{pk(s)} $
&$\alpha.4$.$ \INT(s) \rightarrow a \ : \ s,a,b, v(\emph{r}_1,\emph{re}_1) $
\\

$\beta.4$.&$ s \rightarrow \INT(a) \ : \ s,a,b, v(\emph{r}_1,\emph{r}_2) $ &
\\
$\alpha.4$.&$ \INT(s) \rightarrow a \ : \ s,a,b, v(\emph{r}_1,\emph{r}_2) $
& \\
\hline
\end{tabular}\\[0.7mm]
\end{center}
\end{table}

The attack of Table \ref{es_3} (left side) is successful since the
intruder can manipulate the first three non-encrypted fields. Notice
how $F_2$ is violated: when $a$ receives message $\alpha.4$, $b$
\emph{never} sent message $b,s,a, \{\emph{t}_b, \emph{s}_b,
\emph{r}_2\}_{pk(s)}$.  The attack reported in Table \ref{es_3} (right
side) is stronger, since the principal $b$ is not alive in the run of $a$.

We believe these attacks over this modified version of the TMN
protocol have never been reported before in the literature.
\section{Related Work}
\label{sec:relatedwork}

In this section we discuss some related work.  In \cite{Roscoe96},
Roscoe identifies two ways of specifying protocol security goals:
firstly, using \emph{ extensional} specifications, and secondly using
\emph{ intensional} specifications. An extensional specification
describes the intended service provided by the protocol in terms of
behavioural equivalence \cite{FG94,Aba99,schneider96b}. On
the other hand, an intensional specification describes the underlying
mechanism of a procotol, in terms of states or events
\cite{AB01,WL93,Roscoe96,sm96,Pau98,GT02}.

The approach presented in this paper belongs to the spectrum of
intensional specifications, and is related to \cite{Roscoe96,sm96}.
In \cite{sm96}, a requirement specification language is proposed. This
language is useful for specifying sets of requirements for classes of
protocol; the requirements can be mapped onto a particular protocol
instance, which can be later verified using their tool, called NRL
Protocol Analyzer.  This approach has been subsequently used to
specify
the GDOI secure multicast
protocol \cite{meadowsgdoi}.

In \cite{Roscoe96}, Roscoe presents a method for describing the
underlying mechanism of a protocol, using a CSP specification.  The
method consists of four steps: Firstly, one identifies an execution
point of the protocol that should not be reached without a
corresponding legitimate run having occurred.  Secondly, one describes
the possible sequences of messages that should have occurred before
this execution point; thirdly, one creates a specification which
groups all the CSP processes modelling the protocol participants (this
step is similar to our scenario setting).  Finally, one verifies the
specification using FDR.  This method has been also used by Lowe in
\cite{LR97}.

The approaches just mentioned employ languages specifying security
properties in a global fashion, as opposed to our technique which
deals with local security properties.

In \cite{cmv03}, Cremers, Mauw and de Vink present another logic for
specifying local security properties.  Similarly to us, in
\cite{cmv03} the authors define the message authenticity property
by referring to the variables occurring in the protocol role.
In addition, in \cite{cmv03}, it is defined a new kind of
authentication, called \emph{synchonization}, which is then compared
with the Lowe's intensional specification.  The logic presented in
this paper cannot handle the specification of the synchronization
authentication. In fact, we cannot handle the weaker notion of
injective authentication, since we cannot match corresponding events
in a trace. However, we believe we can extend our logic to support
these properties.  Briefly, this could be achieved by decorating the
different runs with label identifiers and adding a primitive to reason
about events that happenned before others in a trace.


\section{Conclusions and Future Work}
\label{sec:conclusions}
We have developed a \emph{trace logic} for expressing \emph{ local
security properties}.  Using this trace logic, the protocol designer
can specify precisely the (local) security properties a protocol should
satisfy to accomplish the security goals for which it has been
designed.

The main differences between our approach and the ones mentioned in
Section \ref{sec:relatedwork} can be summarized as follows:

\begin{enumerate}

\item Our trace logic formulae are \emph{local} to the participants,
in the sense that are dependent to the principal's point of view,
instead of \emph{global} to the protocol specification. This allow us
to define properties more precisely, in the sense of what should hold
for each principal at each execution step.

Furthermore, our technique can be used to integrate the specification
within the design of a cryptographic protocol. Methodologically, this
allows for the integration of the verification phase within the design
one, speeding up the feedback from the verification, and providing the
basis for an integrated environment for protocol engineering.

\item Without having to use temporal operators, the logic we presented can 
express classical security properties including freshness and
authenticity of the exchanged values during a protocol run.

\item Our logic is applied \emph{directly} to the protocol messages.
This allow us to reason about (local) security properties without
having to use artificial \emph{event} messages.

\end{enumerate}

As future work, we plan to apply the methodology to more complex case
studies, such as multicast protocols e.g. LKH group communications
protocol \cite{LKH}. We also plan to study how to \emph{compose} local
security specifications: we believe this is a very important advantage
of our approach over the other global ones.

\emph{Acknowledgements.}
We would like to thank the anonymous reviewers for useful comments.


\end{document}